\documentclass[prl, aps, twocolumn, groupedaddress,showpacs]{revtex4}
\usepackage{eqnarray}
\usepackage[english]{babel} 
\usepackage[english]{layout}
\usepackage{amsmath,amsfonts,amssymb}
\usepackage{graphicx}
\usepackage{float}
\usepackage{color}
\usepackage{braket}
\usepackage{version}
\usepackage{array,multirow,makecell}

\begin{document}
\title{$\mathbb{Z}_2$ topological insulator analog for vortices in an interacting bosonic quantum fluid}
\author{O. Bleu, G. Malpuech, D. D. Solnyshkov}
\affiliation{Institut Pascal, PHOTON-N2, University Clermont Auvergne, CNRS, 4 avenue Blaise Pascal, 63178 Aubi\`{e}re Cedex, France.} 

\begin{abstract}
$\mathbb{Z}_2$ topological insulators for photons and in general bosons cannot be strictly implemented because of the lack of symmetry-protected pseudospins. We show that the required protection can be provided by the real-space topological excitation of an interacting quantum fluid: quantum vortex. We consider a Bose-Einstein Condensate at the $\Gamma$ point of the Brillouin zone of a quantum valley Hall system based on two staggered honeycomb lattices. We demonstrate the existence of a coupling between the winding number of a vortex and the valley of the bulk Bloch band. This leads to chiral vortex propagation at the zigzag interface between two regions of inverted staggering, where the winding-valley coupling provides true topological protection against backscattering, contrary to the interface states of the non-interacting Hamiltonian. This configuration is an analog of a $\mathbb{Z}_2$ topological insulator for quantum vortices.
\end{abstract}
\maketitle

Topological defects are a distinctive feature of  quantum fluids \cite{LeggettBook}. Such real space excitations are stable and cannot be removed by a continuous transformation, which is called topological protection. They are known for more than fifty years and determine the fluid properties, for example, in the Berezinskii-Kosterlitz-Thouless phase transition in Bose-Einstein Condensates (BECs) \cite{Pitaevskii}.

Since the eighties, the concept of topology has been applied to reciprocal space. The topology of Landau levels \cite{Klitzing1980,TKNN1982,Hatsugai1993} and more generally of Bloch bands \cite{Haldane1988} has been shown to determine the spectacular properties of topological insulators. In this case, the single-particle energy bands of the system are described by topological invariants \cite{Hatsugai1993} (such as the Chern number). The field expanded even further with the discovery of the quantum spin Hall effect and of the associated class of $\mathbb{Z}_2$ topological insulators \cite{Hasan2010,Chiu2016}. Indeed, if one considers spinor particles in a lattice (electrons for instance), the Chern number computed using only one spin component is not a topological invariant. On the other hand, the difference between two spin Chern numbers is a $\mathbb{Z}_2$ topological invariant for a Hamiltonian verifying Time-Reversal Symmetry (TRS) \cite{Kane2005}. In that case the bulk-boundary correspondence applies and guarantees on the interface with a trivial insulator the presence of a pair of counter-propagating spin-polarized states, which because of TRS do not couple the one to the other.

This triumph of topology was followed by the attempts to extend the concept of $\mathbb{Z}_2$ topological insulators to other types of two-level systems which can be mapped to a pseudospin representing either an internal degree of freedom (the polarization of a photon) or an external one (angular momentum states, valleys of a honeycomb lattice \cite{ren2016topological}, etc). However, for photons, TRS acts differently from fermions \cite{Lu2014} and rigorously, there is no symmetry-protected $\mathbb{Z}_2$ photonic topological insulator. This can be clearly visualized by explicitly considering the photonic spin-orbit coupling due to the energy splitting between TE and TM modes \cite{Kavokin2005,Sala2015,CRAS2016}. It respects TRS, but it has a double winding number which couples counter-propagating spin-polarized photonic modes. The realization of a $\mathbb{Z}_2$ topological insulator analog for light therefore requires to fabricate a structure where the TE-TM splitting is weak, which is possible but very demanding \cite{Khanikaev2013,Slobozhanyuk2017}. Other degrees of freedom, like the angular momentum of photons in lattices of ring cavities have been considered \cite{Hafezi2013} with the formal problem that no specific symmetry protects this pseudospin which is affected by disorder. Finally, the quantum valley Hall (QVH) effect in staggered honeycomb lattices uses the valley pseudospin \cite{Niu2007,ren2016topological}. It has been evidenced experimentally in electronic systems \cite{Ju2015} and recently considered in a large series of works in topological photonics \cite{Wu2015,Ma2015,ma2016all,chen2016valley,Xu2016,Barik2016}. Here, the mechanism of dissipation is inter-valley scattering \cite{Bleu2017}. Even if it is argued to be weak in electronic systems and to be zero for certain types of defects respecting the lattice symmetry in photonics, it formally leads to the Anderson localization of the 1D edge states.

The topology of the quantum fluid in real space and of the band in the reciprocal space have already been fruitfully combined in topological superconductors and superfluids \cite{Volovik1999,Lv2017,Sato2017}. The collective excitations of the fluid, described by the Bogoliubov-de Gennes equation, are split off by the superconducting gap, which can become topologically non-trivial for specific shapes of the pairing, creating topological edge states. A vortex, whose core remains in the normal phase, necessarily contains such edge states, which can be Majorana fermions \cite{Elliott2015} protected by the particle-hole symmetry. Many other solitonic \cite{Jackiw76,Takahashi1979,Bartsch2006,Haddad2015,Solnyshkov2016,
Skryabin2017,Solnyshkov2017} and vortex \cite{Haddad2015b} solutions were found in non-trivial topologies, but the chiral behavior has been mostly discussed for weak Bogoliubov excitations \cite{peano2015topological,di2016topological,Bardyn2016,Bleu2016,Gulevich2016,Bleu2017b,
Liew2017}.

In this work, we propose an original combination of real and reciprocal space topologies, creating a truly protected pseudospin current in a bosonic system. 
Here, the topological phase and the edge spin currents are not protected by a symmetry of the Hamiltonian, but by the real-space topology of the quantum vortices. We consider a BEC at the $\Gamma$ point of the Brillouin zone of a QVH system based on two staggered honeycomb lattices. We demonstrate the existence of a coupling between the winding number of a vortex and the valley of the bulk Bloch band. This coupling  leads to chiral vortex propagation at an interface between two regions with inverted staggering, where the winding-valley coupling provides true topological protection against backscattering, contrary to the interface states of the non-interacting Hamiltonian. This configuration can be seen as a $\mathbb{Z}_2$ topological insulator, similar to the quantum spin Hall effect \cite{Kane2005}, but where the role of spin is played by the winding of the vortices. Our results apply to polariton condensates in recently fabricated polariton honeycomb lattices \cite{Jacqmin2014} and to atomic BECs in optical lattices \cite{Soltan2011}.

\paragraph{Non-interacting QVH.}
We consider an interface between two honeycomb lattices with opposite staggering, each of them being well described by a tight-binding (TB) Hamiltonian:
\begin{equation}
H_k=\begin{pmatrix}
\Delta && -Jf_k\\
-Jf_k^* && -\Delta
\end{pmatrix}, ~~ f_k=\sum_{j=1}^3\exp{(-i\textbf{kd}_{\phi_j})} 
\end{equation}
where $2\Delta=E_B-E_A$ is the energy difference between the ground states of A and B sites and $J$ is the nearest neighbour tunnelling coefficient.
A non-zero $\Delta$ leads to the opening of a bandgap and implies the presence of opposite Berry curvatures in $K$ and $K'$ valleys. If the gap is sufficiently small, the Berry curvature is strongly localized in each valley which allows to compute the valley Chern numbers: $C_{K,K'}=\pm 0.5$. The number of chiral states in each valley at the zigzag interface between the opposite lattices is defined by the domain wall topological invariant \cite{Martin2008}: $N_{K,K'}=C_{K,K'}(l)-C_{K,K'}(r)=\pm 1$ (where $l$ and $r$ stand for the left and right domains). This results in the presence of one chiral state in each valley with opposite group velocities (QVH effect). However, these valley states, degenerate in energy, are not protected by some specific symmetry, which means that the backscattering due to diffusion from one valley to the other is not forbidden for single particles.

\paragraph{Quantum vortices.}
The BEC can be described by a single-particle wavefunction (WF) $\psi$ (the order parameter). In the mean-field approximation, $\psi$ is the solution of the Gross-Pitaevskii equation (GPE), including interparticle interactions:
\begin{equation}
i\hbar \frac{{\partial \psi }}{{\partial t}} =  - \frac{{{\hbar ^2}}}{{2m}}\Delta \psi  + \alpha {\left| \psi  \right|^2}\psi +U\psi - \mu \psi 
\label{GPE}
\end{equation}
where $m$ is the particle mass, $\alpha$ is the interaction constant, $U$ is the external potential, and $\mu$ is the chemical potential of the condensate.
The existence of $\psi$ imposes the irrotationality of this bosonic quantum fluid: $\nabla\times \mathbf{v}=0$ everywhere, except zero-density points. The condensate velocity is given by $\mathbf{v}=\hbar\nabla \varphi/m$ ($\varphi=\arg\psi$). The phase winding around the zero-density points where $\psi=0$ is fixed by the single-valuedness of $\psi$: $\oint {\nabla \varphi dl}  = 2\pi p$,
where $p$ is the winding number. The solutions with non-zero $p$ are called vortices, and their characteristic size is determined by the healing length $\xi=\hbar/\sqrt{2\alpha n m}$. We will consider only single-winding vortices ($p=\pm1$), because they are energetically stable. We are going to study such vortex solutions in a staggered honeycomb lattice.

\paragraph{Winding-valley coupling.} 
First, we shall demonstrate that the core of a vortex with a given winding corresponds to a certain valley (K or K') of the single-particle dispersion of staggered graphene, that is, the existence of winding-valley coupling for vortices. 

Let us consider the core of a sufficiently large vortex ($\xi\gg a$, where $a$ is the distance between nearest neighbors), where the density is necessarily small and the interactions can be neglected. To minimize the on-site energy given by $E=E_A |\psi_A|^2+E_B |\psi_B|^2$, the WF is mostly localized on the atoms of the $A$ type, which have lower energy (assuming $E_A<E_B$). In the limit of a large gap, $\Delta\gg J$, only the $A$-atoms are populated.
The WFs in a periodic lattice can be written as a product of a Bloch function and a plane wave.  For the hexagonal lattice, the Bloch part of the WF determines the densities and phases on the $A$ and $B$ atoms.  Therefore the Bloch function in the vicinity of the vortex center is $(1, 0)^T$. We can obtain the corresponding plane wave by Fourier transform of the WF $\widetilde{\psi} \left( {\bf{k}} \right)$ analytically in the TB approximation (see \cite{suppl} for details). We find that the maximum value of the WF is achieved for $k=K$ and $k=K'$, depending on the vortex winding $p$. Thus, both the Bloch wave and the plane wave part of the WF in the core of a vortex of a given winding define a state corresponding to a certain well-defined valley of the single-particle dispersion, and there is a winding-valley coupling for sufficiently large vortices which reads:
\begin{equation}
\tau=ps
\end{equation}
where $\tau=\pm1$ is the valley number and  $s$ is the lattice staggering ($s=+1$ for $E_A<E_B$ and $s=-1$ for $E_A>E_B$). This result is linked with the well-known optical selection rules in Transitional Metal Dichalcogenides \cite{Xiao2012} where the phase pattern at the $K$ point exhibits an angular momentum for \emph{each} unit cell, which  determines the angular momentum of photons absorbed for a given valley.

\begin{figure}[tbp]
 \begin{center}
 \includegraphics[scale=0.35]{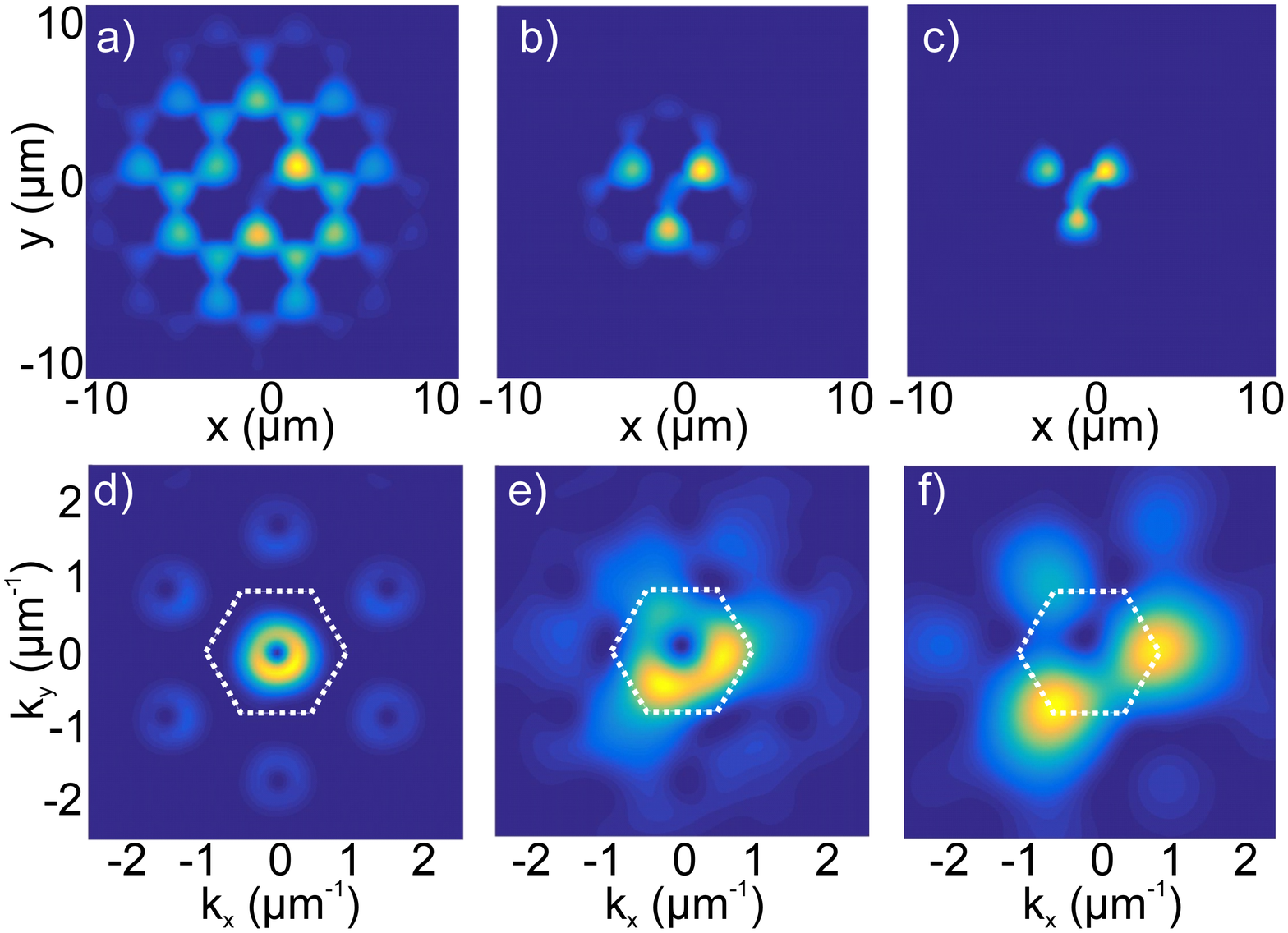}
 \caption{\label{coreNum} Numerical density profile of the vortex stationary solution in real(a,b,c) and reciprocal (d,e,f) space for different filtering scales ($w=7,3,1$~$\mu$m, respectively). All parameters as in Ref. \cite{Bleu2017}.}
 \end{center}
 \end{figure}

To confirm our analytical TB solution, we have performed numerical simulations by solving the GPE beyond the TB approximation, with an explicit honeycomb lattice potential $U(r)$. To find the WF of the vortex, we have introduced the relaxation term \cite{Pitaevskii58}, preserving zeroes of the WF. 
The results of these calculations are shown in Fig. \ref{coreNum}. We filter the vortex core using a Gaussian function of size $w$ (panels (a)-(c)). For large $w$, the image in the reciprocal space (Fig.~\ref{coreNum}(d)) is dominated by the condensate centered at the ground state ($\Gamma$ point). The ground state itself is empty, because the vortex imposes $v\neq 0$ everywhere. For smaller $w$ (Fig.~\ref{coreNum}(e,f)), the core of the vortex is centered at the $K$ points of the reciprocal space, while the $K'$ valleys are empty. Opposite results are obtained for opposite winding, confirming the valley-winding coupling for vortices.

\paragraph{Vortex at the interface.}
We have shown that the vortex WF in the reciprocal space is composed of 2 important contributions. Most of the particles of the condensate, far from the vortex core, are concentrated around the $\Gamma$ point (small $k$). These particles are practically unaffected neither by the presence of the lattice, nor by any possible interfaces.
On the other hand, the core of the vortex is at the $K$ point, and we can expect interesting effects linked with the interfaces, where in the linear regime the states from the bulk $K$ points give rise to chiral propagative interface states (QVH states). We shall therefore calculate
analytically the energy of the vortex as a function of its position and wavevector of the core, using the TB approximation.

In a general case, the energy of the vortex can be calculated using the grand canonical expression \cite{Pitaevskii}:
\begin{equation}
{E_v} = \int {\left( {\frac{{{\hbar ^2}}}{{2m}}{{\left| {\nabla \psi } \right|}^2} + \frac{\alpha }{2}{{\left( {{{\left| \psi  \right|}^2} - n} \right)}^2}} \right)\mathbf{dR}} 
\end{equation}
Qualitatively, this expression is the difference between the energy of a system with a vortex and the energy of a system without a vortex (but with a condensate in the ground state with the unperturbed density $n$). The first step is to split the integral into 2 regions: the core ($|\mathbf{R}|\leq\xi$) and the outside zone ($|\mathbf{R}|>\xi$). In the second region, $|\psi|^2\approx n$, and the only contribution to the vortex energy comes from the kinetic energy term, which gives the well-known logarithmic expression \footnote{This result does not depend on the presence of a lattice, because it corresponds to large distances and small wavevectors (long-wavelength approximation). It does not depend on the vortex position, because the local potential does not affect the overall rotation of the particles dominating this energy. Neither does it depend on the propagation of the vortex, because the condensate far from the core remains globally unperturbed by this propagation by definition, otherwise the situation would correspond to the propagation of the vortex with a flow.} $
E_v^{r > \xi } = \pi n\hbar^2\ln \left( 1.46 R_0/\xi\right)/m$ ($R_0$ is the system size).

\begin{figure}[tbp]
 \includegraphics[scale=0.3]{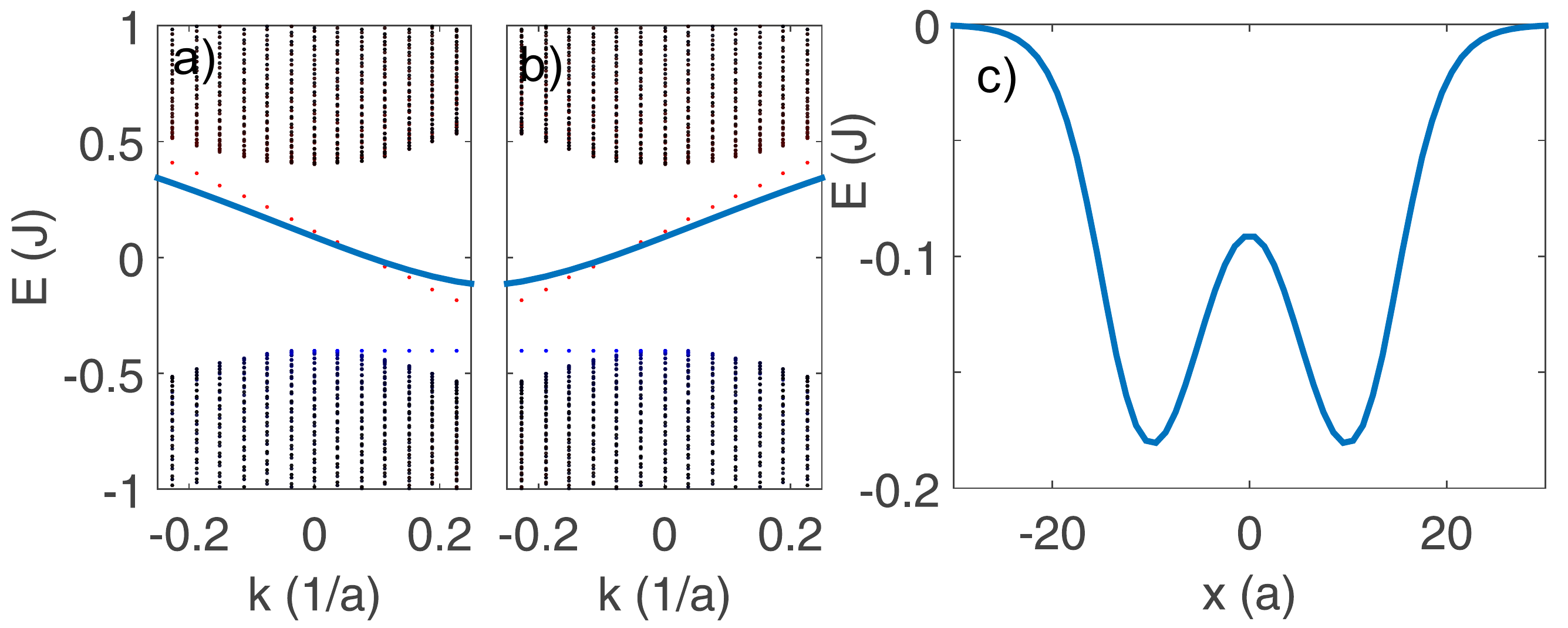}
 \caption{\label{OlivierFigure} a,b) Energy of the vortex core at the interface as a function of its central wavevector, exhibiting valley chirality (a - $K$, b - $K'$). c) Energy of the vortex as a function of position.}
 \end{figure}

In the vortex core, the presence of the lattice has to be taken into account. As we have shown above both analytically and numerically, the core of the vortex is a wavepacket centered at a wavevector $k_0$ close to either $K$ or $K'$ (we take a Gaussian wavepacket $\psi_G$). We calculate its energy versus $k_0$ using the TB dispersion $E(k)$. 
 However,  the $X$ spatial direction, perpendicular to the interface, has to be treated in the real space ($x_0$ is the vortex center). The contribution to the kinetic energy is calculated as: $
E_v^{kin,r<\xi}(x_0,k_0)=\int_{x_0-\xi}^{x_0+\xi} dx \int dk \psi_G^*\psi^*_0\hat{H}\psi_0\psi_G$,
where $\psi_0(x,k)$ are the single-particle eigenstates of the lattice. These eigenstates are quantized in the $X$ direction. Their spatial overlap with the vortex core plays an important role. For the delocalized bulk states the overlap tends to zero with increase of the stripe width. On the other hand, the state localized at the interface (width $\kappa$) has a non-vanishing overlap and the contribution of this state dominates the dispersion of the vortex core. An example of such dispersion in the vicinity of the $K$ and $K'$ points is shown in Fig. \ref{OlivierFigure}(a,b): the dispersion of the core (blue line) inherits the dispersion of the linear eigenstates at the interface (red dots), and therefore their valley-dependent propagation direction (chirality), as compared with the non-propagating bulk states with zero group velocity exactly at $K$ or $K'$ (black points). 

The kinetic energy of the core also depends on the position of its center $x_0$: if the core is perfectly superposed with the interface state (centered at the interface), the energy at $k_0=K$ is exactly the same as that of the interface state. On the other hand, if the core is located in the bulk, its energy is that of the top of the valence band, determined by the energy splitting $E^{kin}(x_0,k_0)=-\Delta$. The interface therefore represents a \emph{barrier}, if only the kinetic energy is taken into account.

The contribution of the interactions to the vortex core comes from the sensitivity of the vortex to the local changes of the density in the condensate. In the vortex core, the density $|\psi|^2$ is small as compared with the background density $n(\mathbf{r})$, and the integral reads: $
E_v^{int,r < \xi } = \int_0^\xi  \alpha {n^2}\pi rdr
$. Thus, the vortices are attracted to lower-density regions minimizing the total energy of the system. The density of the condensate without a vortex depends on the local potential, which affects the density of the condensate at the scale given by the healing length $\xi$. Considering the interface as a Delta barrier $V_0\delta(x)$, the density of the condensate in its presence can be found as \cite{Tanese2013}:
$n(x)=n_0(1-\cosh^2((x_c+|x|)/\xi'))$,
where $x_c$ and $\xi'$ depend on $V_0$. The interaction energy of the vortex core as a function of $x_0$ therefore exhibits a \emph{minimum} of the width $\xi'\approx\xi$.

The sum of the kinetic and interaction energy depends on the parameters of the system. An example of such dependence as a function of $x_0$ is shown in Fig. \ref{OlivierFigure}(c) for $\xi>\kappa$. In this case, the vortex can be localized on either side of the interface, the latter acting as a barrier  preventing the vortex to go to the other side of the interface and change valley. We see that the properties of the single-particle dispersion of the interface states are inherited by the vortex solution of the non-linear equation via the core.

\begin{figure}[tbp]
 \begin{center}
 \includegraphics[scale=0.63]{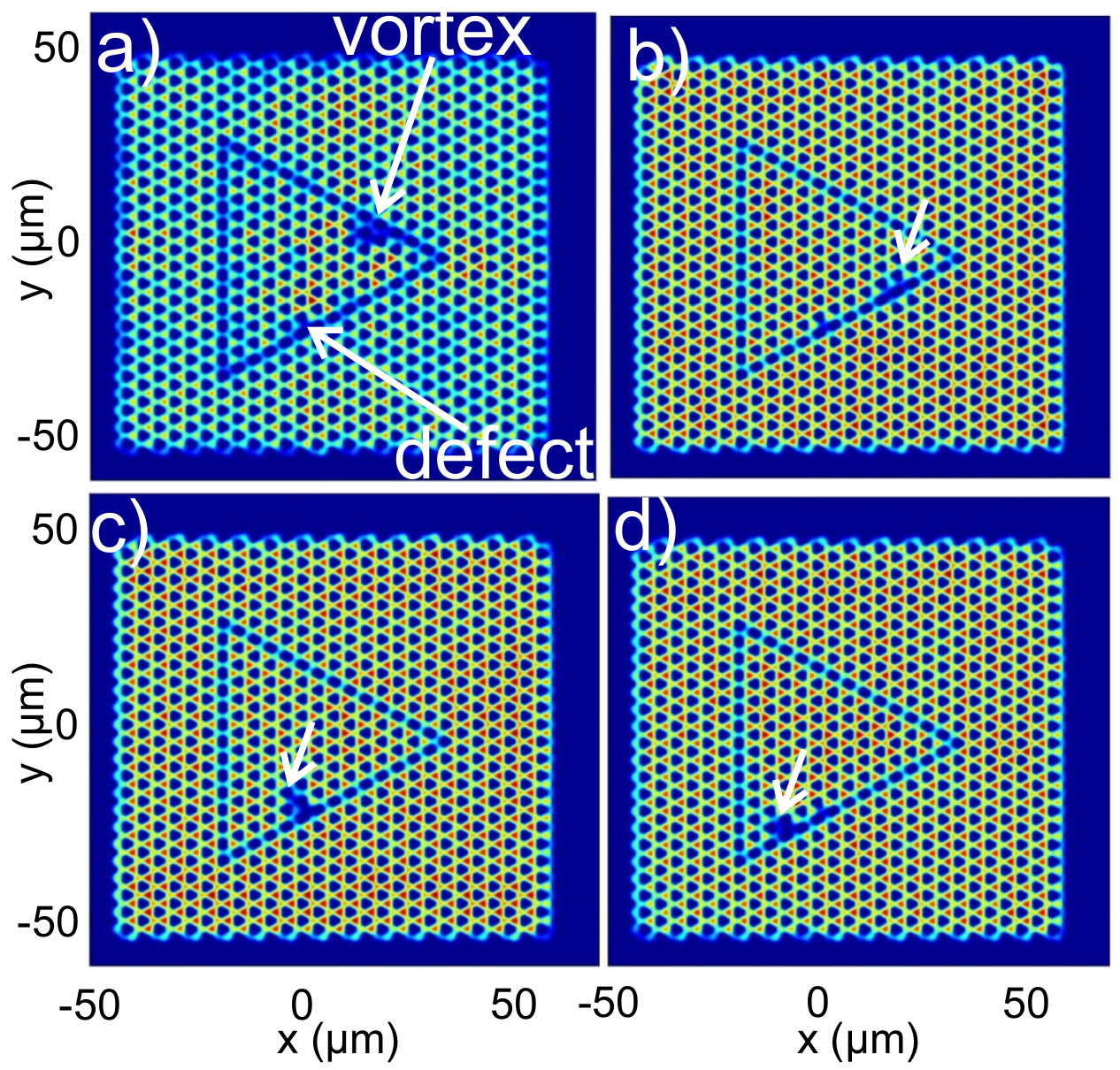}
 \caption{\label{snap} Snapshots of the vortex propagation along the interface, showing the spatial density distribution $|\psi(x,y)|^2$.}
  \end{center}
 \end{figure}

Our analytical results are again fully confirmed by numerical simulations of vortex propagation along the interface using Eq.~\eqref{GPE}. The snapshots of one of such simulations are shown in Fig. \ref{snap} (see \cite{suppl} for movies). We see that the vortex remains attached to the interface and propagates along, without being scattered backwards on the corners. An additional defect of 1 meV has been added on an interface pillar for comparison with the linear case, where it led to strong backscattering \cite{Bleu2017}, 
which allows us to check that the vortex is indeed immune to backscattering thanks to the additional topological protection provided by its winding via the winding-valley coupling.
For direct comparison, all parameters used in numerical simulations were exactly the same as in \cite{Bleu2017} (except interactions \cite{suppl}). It allows to obtain the group velocity of the interface states $\hbar v_g=\partial E/\partial k=0.7\times 10^6$ m/s or $0.7$ $\mu$m/ps. This is the velocity with which the WPs at the interface can be expected to propagate in this particular lattice. Interestingly, the vortex velocity is different from $v_g$. We stress that it is also different from what can be calculated for the vortex rolling effect (see \cite{suppl}).

\begin{figure}[tbp]
 \begin{center}
 \includegraphics[scale=0.43]{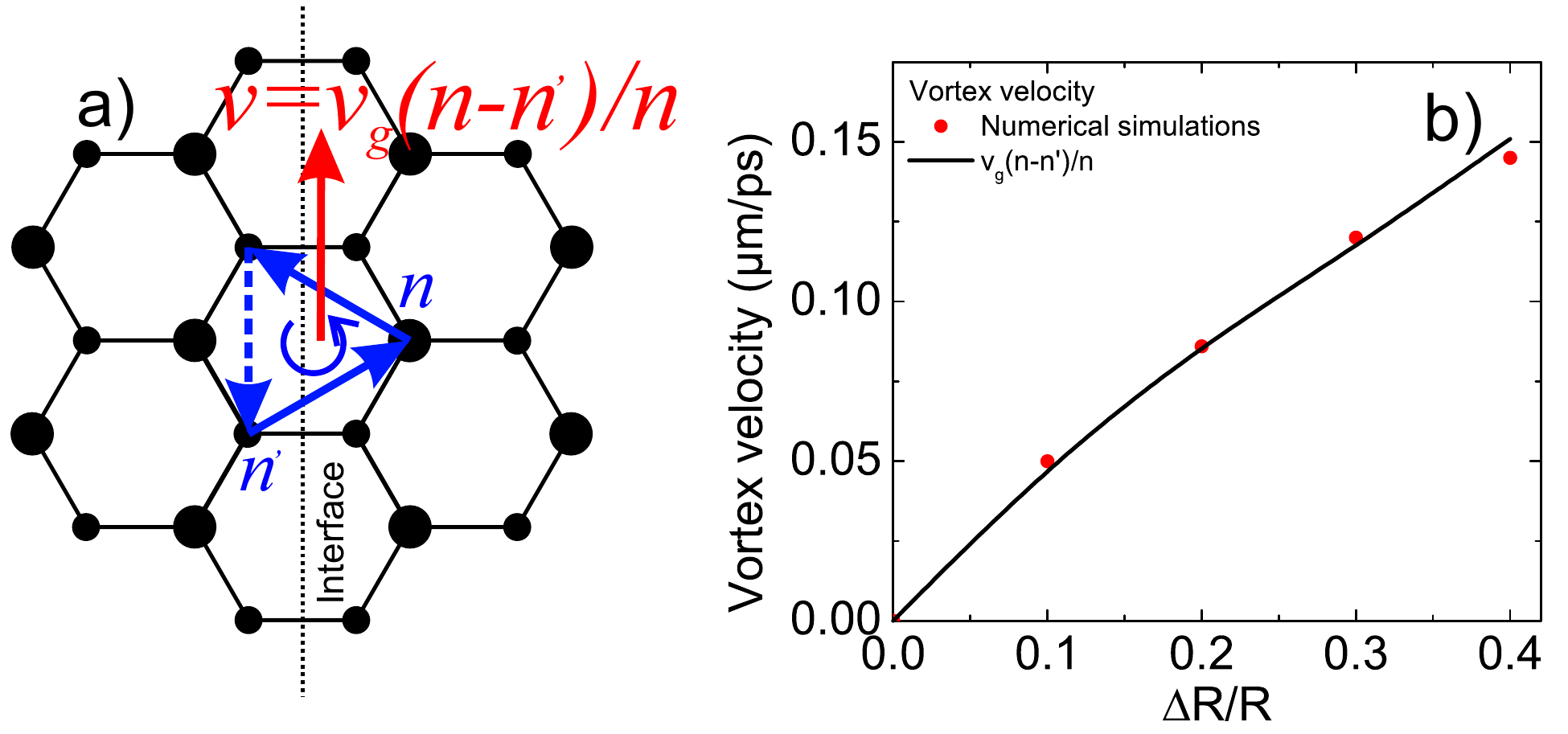}
 \caption{\label{figVscheme} a) A vortex at an interface and its net velocity.  b) Vortex velocity as a function of the gap size. Red points - numerical results, black - analytical solution.}
  \end{center}
 \end{figure} 

Indeed, in our calculation we were assuming that only one type of the atoms is occupied for a given staggering. However, as shown in a scheme in Fig.~\ref{figVscheme}(a), the interface represents a violation of a perfect staggering, and thus the higher-energy sublattice acquires a density  estimated as  $n' = 2n/\left(1 + \left(\Delta  + \sqrt {\Delta ^2 + 4J^2}  \right)^2/4{J^2}\right)$ (see \cite{suppl}). 
 The resulting velocity, reduced with respect to that of the linear interface states, is given by:
\begin{equation}
v=v_g\left(n-n'\right)/n
\label{gvel}
\end{equation}
We plot the dependence of $v$ on the pillar size ratio $\Delta R/R$ (determining the gap size $\Delta$) in  Fig.~\ref{figVscheme}(b). Red dots show the results of numerical simulations. Black line is the analytical solution given by Eq.~\eqref{gvel}, where $v_g$ and $\Delta$ are taken from numerical simulations in linear regime. We see that it corresponds almost perfectly to the points (exact numerical solution) while there are no fitting parameters. This confirms the validity of our interpretation.

\emph{Conclusions.} 
We demonstrate a vortex-valley coupling for a BEC in a staggered honeycomb potential. The main consequence of this property is the robust chiral propagation of vortices at the zigzag interface between two lattices with opposite staggering. The vortices, contrary to the linear WPs, are immune from backscattering thanks to their real-space topological protection. Hence, this work highlights a new combination of real and momentum space topology.  These results are promising for the development of a new field of vortextronics, where the information will be carried by vortices. The possibility to create chiral pathways for vortices and to automatically sort them according to their winding is crucial for information treatment.

\begin{acknowledgments}
We acknowledge the support of the project "Quantum Fluids of Light"  (ANR-16-CE30-0021), of the ANR Labex Ganex (ANR-11-LABX-0014), and of the ANR Labex IMobS3 (ANR-10-LABX-16-01). D.D.S. acknowledges the support of IUF (Institut Universitaire de France).
\end{acknowledgments}

\bibliography{biblio} 

\begin{thebibliography}{52}
\expandafter\ifx\csname natexlab\endcsname\relax\def\natexlab#1{#1}\fi
\expandafter\ifx\csname bibnamefont\endcsname\relax
  \def\bibnamefont#1{#1}\fi
\expandafter\ifx\csname bibfnamefont\endcsname\relax
  \def\bibfnamefont#1{#1}\fi
\expandafter\ifx\csname citenamefont\endcsname\relax
  \def\citenamefont#1{#1}\fi
\expandafter\ifx\csname url\endcsname\relax
  \def\url#1{\texttt{#1}}\fi
\expandafter\ifx\csname urlprefix\endcsname\relax\def\urlprefix{URL }\fi
\providecommand{\bibinfo}[2]{#2}
\providecommand{\eprint}[2][]{\url{#2}}

\bibitem[{\citenamefont{Leggett}(2006)}]{LeggettBook}
\bibinfo{author}{\bibfnamefont{A.}~\bibnamefont{Leggett}},
  \emph{\bibinfo{title}{Quantum Liquids}} (\bibinfo{publisher}{Oxford Graduate
  Texts}, \bibinfo{year}{2006}).

\bibitem[{\citenamefont{Pitaevskii and Stringari}(2003)}]{Pitaevskii}
\bibinfo{author}{\bibfnamefont{L.}~\bibnamefont{Pitaevskii}} \bibnamefont{and}
  \bibinfo{author}{\bibfnamefont{S.}~\bibnamefont{Stringari}},
  \emph{\bibinfo{title}{Bose-Einstein Condensation}}
  (\bibinfo{publisher}{Oxford Science Publications - International Series of
  Monographs on Physics 116}, \bibinfo{year}{2003}).

\bibitem[{\citenamefont{Klitzing et~al.}(1980)\citenamefont{Klitzing, Dorda,
  and Pepper}}]{Klitzing1980}
\bibinfo{author}{\bibfnamefont{K.~v.} \bibnamefont{Klitzing}},
  \bibinfo{author}{\bibfnamefont{G.}~\bibnamefont{Dorda}}, \bibnamefont{and}
  \bibinfo{author}{\bibfnamefont{M.}~\bibnamefont{Pepper}},
  \bibinfo{journal}{Phys. Rev. Lett.} \textbf{\bibinfo{volume}{45}},
  \bibinfo{pages}{494} (\bibinfo{year}{1980}).

\bibitem[{\citenamefont{Thouless et~al.}(1982)\citenamefont{Thouless, Kohmoto,
  Nightingale, and den Nijs}}]{TKNN1982}
\bibinfo{author}{\bibfnamefont{D.~J.} \bibnamefont{Thouless}},
  \bibinfo{author}{\bibfnamefont{M.}~\bibnamefont{Kohmoto}},
  \bibinfo{author}{\bibfnamefont{M.~P.} \bibnamefont{Nightingale}},
  \bibnamefont{and} \bibinfo{author}{\bibfnamefont{M.}~\bibnamefont{den Nijs}},
  \bibinfo{journal}{Phys. Rev. Lett.} \textbf{\bibinfo{volume}{49}},
  \bibinfo{pages}{405} (\bibinfo{year}{1982}).

\bibitem[{\citenamefont{Hatsugai}(1993)}]{Hatsugai1993}
\bibinfo{author}{\bibfnamefont{Y.}~\bibnamefont{Hatsugai}},
  \bibinfo{journal}{Phys. Rev. Lett.} \textbf{\bibinfo{volume}{71}},
  \bibinfo{pages}{3697} (\bibinfo{year}{1993}),
  \urlprefix\url{http://link.aps.org/doi/10.1103/PhysRevLett.71.3697}.

\bibitem[{\citenamefont{Haldane}(1988)}]{Haldane1988}
\bibinfo{author}{\bibfnamefont{F.~D.~M.} \bibnamefont{Haldane}},
  \bibinfo{journal}{Phys. Rev. Lett.} \textbf{\bibinfo{volume}{61}},
  \bibinfo{pages}{2015} (\bibinfo{year}{1988}),
  \urlprefix\url{https://link.aps.org/doi/10.1103/PhysRevLett.61.2015}.

\bibitem[{\citenamefont{Hasan and Kane}(2010)}]{Hasan2010}
\bibinfo{author}{\bibfnamefont{M.~Z.} \bibnamefont{Hasan}} \bibnamefont{and}
  \bibinfo{author}{\bibfnamefont{C.~L.} \bibnamefont{Kane}},
  \bibinfo{journal}{Rev. Mod. Phys.} \textbf{\bibinfo{volume}{82}},
  \bibinfo{pages}{3045} (\bibinfo{year}{2010}),
  \urlprefix\url{http://link.aps.org/doi/10.1103/RevModPhys.82.3045}.

\bibitem[{\citenamefont{Chiu et~al.}(2016)\citenamefont{Chiu, Teo, Schnyder,
  and Ryu}}]{Chiu2016}
\bibinfo{author}{\bibfnamefont{C.-K.} \bibnamefont{Chiu}},
  \bibinfo{author}{\bibfnamefont{J.~C.~Y.} \bibnamefont{Teo}},
  \bibinfo{author}{\bibfnamefont{A.~P.} \bibnamefont{Schnyder}},
  \bibnamefont{and} \bibinfo{author}{\bibfnamefont{S.}~\bibnamefont{Ryu}},
  \bibinfo{journal}{Rev. Mod. Phys.} \textbf{\bibinfo{volume}{88}},
  \bibinfo{pages}{035005} (\bibinfo{year}{2016}),
  \urlprefix\url{https://link.aps.org/doi/10.1103/RevModPhys.88.035005}.

\bibitem[{\citenamefont{Kane and Mele}(2005)}]{Kane2005}
\bibinfo{author}{\bibfnamefont{C.~L.} \bibnamefont{Kane}} \bibnamefont{and}
  \bibinfo{author}{\bibfnamefont{E.~J.} \bibnamefont{Mele}},
  \bibinfo{journal}{Phys. Rev. Lett.} \textbf{\bibinfo{volume}{95}},
  \bibinfo{pages}{146802} (\bibinfo{year}{2005}),
  \urlprefix\url{https://link.aps.org/doi/10.1103/PhysRevLett.95.146802}.

\bibitem[{\citenamefont{Ren et~al.}(2016)\citenamefont{Ren, Qiao, and
  Niu}}]{ren2016topological}
\bibinfo{author}{\bibfnamefont{Y.}~\bibnamefont{Ren}},
  \bibinfo{author}{\bibfnamefont{Z.}~\bibnamefont{Qiao}}, \bibnamefont{and}
  \bibinfo{author}{\bibfnamefont{Q.}~\bibnamefont{Niu}},
  \bibinfo{journal}{Reports on Progress in Physics}
  \textbf{\bibinfo{volume}{79}}, \bibinfo{pages}{066501}
  (\bibinfo{year}{2016}).

\bibitem[{\citenamefont{Lu et~al.}(2014)\citenamefont{Lu, Joannopoulos, and
  Soljacic}}]{Lu2014}
\bibinfo{author}{\bibfnamefont{L.}~\bibnamefont{Lu}},
  \bibinfo{author}{\bibfnamefont{J.~D.} \bibnamefont{Joannopoulos}},
  \bibnamefont{and} \bibinfo{author}{\bibfnamefont{M.}~\bibnamefont{Soljacic}},
  \bibinfo{journal}{Nature Photonics} \textbf{\bibinfo{volume}{8}},
  \bibinfo{pages}{821} (\bibinfo{year}{2014}).

\bibitem[{\citenamefont{Kavokin et~al.}(2005)\citenamefont{Kavokin, Malpuech,
  and Glazov}}]{Kavokin2005}
\bibinfo{author}{\bibfnamefont{A.}~\bibnamefont{Kavokin}},
  \bibinfo{author}{\bibfnamefont{G.}~\bibnamefont{Malpuech}}, \bibnamefont{and}
  \bibinfo{author}{\bibfnamefont{M.}~\bibnamefont{Glazov}},
  \bibinfo{journal}{Phys. Rev. Lett.} \textbf{\bibinfo{volume}{95}},
  \bibinfo{pages}{136601} (\bibinfo{year}{2005}),
  \urlprefix\url{http://link.aps.org/doi/10.1103/PhysRevLett.95.136601}.

\bibitem[{\citenamefont{Sala et~al.}(2015)\citenamefont{Sala, Solnyshkov,
  Carusotto, Jacqmin, Lema\^{\i}tre, Ter\ifmmode~\mbox{\c{c}}\else
  \c{c}\fi{}as, Nalitov, Abbarchi, Galopin, Sagnes et~al.}}]{Sala2015}
\bibinfo{author}{\bibfnamefont{V.~G.} \bibnamefont{Sala}},
  \bibinfo{author}{\bibfnamefont{D.~D.} \bibnamefont{Solnyshkov}},
  \bibinfo{author}{\bibfnamefont{I.}~\bibnamefont{Carusotto}},
  \bibinfo{author}{\bibfnamefont{T.}~\bibnamefont{Jacqmin}},
  \bibinfo{author}{\bibfnamefont{A.}~\bibnamefont{Lema\^{\i}tre}},
  \bibinfo{author}{\bibfnamefont{H.}~\bibnamefont{Ter\ifmmode~\mbox{\c{c}}\else
  \c{c}\fi{}as}}, \bibinfo{author}{\bibfnamefont{A.}~\bibnamefont{Nalitov}},
  \bibinfo{author}{\bibfnamefont{M.}~\bibnamefont{Abbarchi}},
  \bibinfo{author}{\bibfnamefont{E.}~\bibnamefont{Galopin}},
  \bibinfo{author}{\bibfnamefont{I.}~\bibnamefont{Sagnes}},
  \bibnamefont{et~al.}, \bibinfo{journal}{Phys. Rev. X}
  \textbf{\bibinfo{volume}{5}}, \bibinfo{pages}{011034} (\bibinfo{year}{2015}),
  \urlprefix\url{https://link.aps.org/doi/10.1103/PhysRevX.5.011034}.

\bibitem[{\citenamefont{Solnyshkov and Malpuech}(2016)}]{CRAS2016}
\bibinfo{author}{\bibfnamefont{D.}~\bibnamefont{Solnyshkov}} \bibnamefont{and}
  \bibinfo{author}{\bibfnamefont{G.}~\bibnamefont{Malpuech}},
  \bibinfo{journal}{Comptes Rendus Physique} \textbf{\bibinfo{volume}{17}},
  \bibinfo{pages}{920 } (\bibinfo{year}{2016}), ISSN \bibinfo{issn}{1631-0705},
  \bibinfo{note}{polariton physics / Physique des polaritons},
  \urlprefix\url{http://www.sciencedirect.com/science/article/pii/S1631070516300500}.

\bibitem[{\citenamefont{Khanikaev et~al.}(2013)\citenamefont{Khanikaev,
  Mousavi, Tse, Kargarian, MacDonald, and Shvets}}]{Khanikaev2013}
\bibinfo{author}{\bibfnamefont{A.~B.} \bibnamefont{Khanikaev}},
  \bibinfo{author}{\bibfnamefont{S.~H.} \bibnamefont{Mousavi}},
  \bibinfo{author}{\bibfnamefont{W.-K.} \bibnamefont{Tse}},
  \bibinfo{author}{\bibfnamefont{M.}~\bibnamefont{Kargarian}},
  \bibinfo{author}{\bibfnamefont{A.~H.} \bibnamefont{MacDonald}},
  \bibnamefont{and} \bibinfo{author}{\bibfnamefont{G.}~\bibnamefont{Shvets}},
  \bibinfo{journal}{Nature Materials} \textbf{\bibinfo{volume}{12}},
  \bibinfo{pages}{233} (\bibinfo{year}{2013}).

\bibitem[{\citenamefont{Slobozhanyuk et~al.}(2017)\citenamefont{Slobozhanyuk,
  Mousavi, Ni, Smirnova, Kivshar, and Khanikaev}}]{Slobozhanyuk2017}
\bibinfo{author}{\bibfnamefont{A.}~\bibnamefont{Slobozhanyuk}},
  \bibinfo{author}{\bibfnamefont{S.~H.} \bibnamefont{Mousavi}},
  \bibinfo{author}{\bibfnamefont{X.}~\bibnamefont{Ni}},
  \bibinfo{author}{\bibfnamefont{D.}~\bibnamefont{Smirnova}},
  \bibinfo{author}{\bibfnamefont{Y.~S.} \bibnamefont{Kivshar}},
  \bibnamefont{and} \bibinfo{author}{\bibfnamefont{A.~B.}
  \bibnamefont{Khanikaev}}, \bibinfo{journal}{Nature Photonics}
  \textbf{\bibinfo{volume}{11}}, \bibinfo{pages}{130} (\bibinfo{year}{2017}).

\bibitem[{\citenamefont{Hafezi et~al.}(2013)\citenamefont{Hafezi, Mittal, Fan,
  Migdall, and Taylor}}]{Hafezi2013}
\bibinfo{author}{\bibfnamefont{M.}~\bibnamefont{Hafezi}},
  \bibinfo{author}{\bibfnamefont{S.}~\bibnamefont{Mittal}},
  \bibinfo{author}{\bibfnamefont{J.}~\bibnamefont{Fan}},
  \bibinfo{author}{\bibfnamefont{A.}~\bibnamefont{Migdall}}, \bibnamefont{and}
  \bibinfo{author}{\bibfnamefont{J.~M.} \bibnamefont{Taylor}},
  \bibinfo{journal}{Nature Photonics} \textbf{\bibinfo{volume}{7}},
  \bibinfo{pages}{1001} (\bibinfo{year}{2013}).

\bibitem[{\citenamefont{Xiao et~al.}(2007)\citenamefont{Xiao, Yao, and
  Niu}}]{Niu2007}
\bibinfo{author}{\bibfnamefont{D.}~\bibnamefont{Xiao}},
  \bibinfo{author}{\bibfnamefont{W.}~\bibnamefont{Yao}}, \bibnamefont{and}
  \bibinfo{author}{\bibfnamefont{Q.}~\bibnamefont{Niu}},
  \bibinfo{journal}{Phys. Rev. Lett.} \textbf{\bibinfo{volume}{99}},
  \bibinfo{pages}{236809} (\bibinfo{year}{2007}),
  \urlprefix\url{http://link.aps.org/doi/10.1103/PhysRevLett.99.236809}.

\bibitem[{\citenamefont{Ju et~al.}(2015)\citenamefont{Ju, Shi, Nair, Lv,
  Velasco, Ojeda-Aristizabal, Bechtel, Martin, Zettl, Analytis
  et~al.}}]{Ju2015}
\bibinfo{author}{\bibfnamefont{L.}~\bibnamefont{Ju}},
  \bibinfo{author}{\bibfnamefont{Z.}~\bibnamefont{Shi}},
  \bibinfo{author}{\bibfnamefont{N.}~\bibnamefont{Nair}},
  \bibinfo{author}{\bibfnamefont{Y.}~\bibnamefont{Lv}},
  \bibinfo{author}{\bibfnamefont{J.~J.} \bibnamefont{Velasco}},
  \bibinfo{author}{\bibfnamefont{C.}~\bibnamefont{Ojeda-Aristizabal}},
  \bibinfo{author}{\bibfnamefont{H.~A.} \bibnamefont{Bechtel}},
  \bibinfo{author}{\bibfnamefont{M.~C.} \bibnamefont{Martin}},
  \bibinfo{author}{\bibfnamefont{A.}~\bibnamefont{Zettl}},
  \bibinfo{author}{\bibfnamefont{J.}~\bibnamefont{Analytis}},
  \bibnamefont{et~al.}, \bibinfo{journal}{Nature}
  \textbf{\bibinfo{volume}{520}}, \bibinfo{pages}{650} (\bibinfo{year}{2015}).

\bibitem[{\citenamefont{Wu and Hu}(2015)}]{Wu2015}
\bibinfo{author}{\bibfnamefont{L.-H.} \bibnamefont{Wu}} \bibnamefont{and}
  \bibinfo{author}{\bibfnamefont{X.}~\bibnamefont{Hu}}, \bibinfo{journal}{Phys.
  Rev. Lett.} \textbf{\bibinfo{volume}{114}}, \bibinfo{pages}{223901}
  (\bibinfo{year}{2015}),
  \urlprefix\url{https://link.aps.org/doi/10.1103/PhysRevLett.114.223901}.

\bibitem[{\citenamefont{Ma et~al.}(2015)\citenamefont{Ma, Khanikaev, Mousavi,
  and Shvets}}]{Ma2015}
\bibinfo{author}{\bibfnamefont{T.}~\bibnamefont{Ma}},
  \bibinfo{author}{\bibfnamefont{A.~B.} \bibnamefont{Khanikaev}},
  \bibinfo{author}{\bibfnamefont{S.~H.} \bibnamefont{Mousavi}},
  \bibnamefont{and} \bibinfo{author}{\bibfnamefont{G.}~\bibnamefont{Shvets}},
  \bibinfo{journal}{Phys. Rev. Lett.} \textbf{\bibinfo{volume}{114}},
  \bibinfo{pages}{127401} (\bibinfo{year}{2015}),
  \urlprefix\url{https://link.aps.org/doi/10.1103/PhysRevLett.114.127401}.

\bibitem[{\citenamefont{Ma and Shvets}(2016)}]{ma2016all}
\bibinfo{author}{\bibfnamefont{T.}~\bibnamefont{Ma}} \bibnamefont{and}
  \bibinfo{author}{\bibfnamefont{G.}~\bibnamefont{Shvets}},
  \bibinfo{journal}{New Journal of Physics} \textbf{\bibinfo{volume}{18}},
  \bibinfo{pages}{025012} (\bibinfo{year}{2016}).

\bibitem[{\citenamefont{Chen and Dong}(2016)}]{chen2016valley}
\bibinfo{author}{\bibfnamefont{X.-D.} \bibnamefont{Chen}} \bibnamefont{and}
  \bibinfo{author}{\bibfnamefont{J.-W.} \bibnamefont{Dong}},
  \bibinfo{journal}{arXiv:1602.03352}  (\bibinfo{year}{2016}).

\bibitem[{\citenamefont{Xu et~al.}(2016)\citenamefont{Xu, Wang, Xu, Chen, and
  Jiang}}]{Xu2016}
\bibinfo{author}{\bibfnamefont{L.}~\bibnamefont{Xu}},
  \bibinfo{author}{\bibfnamefont{H.-X.} \bibnamefont{Wang}},
  \bibinfo{author}{\bibfnamefont{Y.-D.} \bibnamefont{Xu}},
  \bibinfo{author}{\bibfnamefont{H.-Y.} \bibnamefont{Chen}}, \bibnamefont{and}
  \bibinfo{author}{\bibfnamefont{J.-H.} \bibnamefont{Jiang}},
  \bibinfo{journal}{Opt. Express} \textbf{\bibinfo{volume}{24}},
  \bibinfo{pages}{18059} (\bibinfo{year}{2016}),
  \urlprefix\url{http://www.opticsexpress.org/abstract.cfm?URI=oe-24-16-18059}.

\bibitem[{\citenamefont{Barik et~al.}(2016)\citenamefont{Barik, Miyake,
  DeGottardi, Waks, and Hafezi}}]{Barik2016}
\bibinfo{author}{\bibfnamefont{S.}~\bibnamefont{Barik}},
  \bibinfo{author}{\bibfnamefont{H.}~\bibnamefont{Miyake}},
  \bibinfo{author}{\bibfnamefont{W.}~\bibnamefont{DeGottardi}},
  \bibinfo{author}{\bibfnamefont{E.}~\bibnamefont{Waks}}, \bibnamefont{and}
  \bibinfo{author}{\bibfnamefont{M.}~\bibnamefont{Hafezi}},
  \bibinfo{journal}{New Journal of Physics} \textbf{\bibinfo{volume}{18}},
  \bibinfo{pages}{113013} (\bibinfo{year}{2016}),
  \urlprefix\url{http://stacks.iop.org/1367-2630/18/i=11/a=113013}.

\bibitem[{\citenamefont{Bleu et~al.}(2017{\natexlab{a}})\citenamefont{Bleu,
  Solnyshkov, and Malpuech}}]{Bleu2017}
\bibinfo{author}{\bibfnamefont{O.}~\bibnamefont{Bleu}},
  \bibinfo{author}{\bibfnamefont{D.~D.} \bibnamefont{Solnyshkov}},
  \bibnamefont{and} \bibinfo{author}{\bibfnamefont{G.}~\bibnamefont{Malpuech}},
  \bibinfo{journal}{Phys. Rev. B} \textbf{\bibinfo{volume}{95}},
  \bibinfo{pages}{235431} (\bibinfo{year}{2017}{\natexlab{a}}),
  \urlprefix\url{https://link.aps.org/doi/10.1103/PhysRevB.95.235431}.

\bibitem[{\citenamefont{Volovik}(1999)}]{Volovik1999}
\bibinfo{author}{\bibfnamefont{G.~E.} \bibnamefont{Volovik}},
  \bibinfo{journal}{JETP Lett.} \textbf{\bibinfo{volume}{70}},
  \bibinfo{pages}{609} (\bibinfo{year}{1999}).

\bibitem[{\citenamefont{Lv et~al.}(2017)\citenamefont{Lv, Wang, Zhang, Ding,
  Li, Wang, He, Song, Ma, and Xue}}]{Lv2017}
\bibinfo{author}{\bibfnamefont{Y.-F.} \bibnamefont{Lv}},
  \bibinfo{author}{\bibfnamefont{W.-L.} \bibnamefont{Wang}},
  \bibinfo{author}{\bibfnamefont{Y.-M.} \bibnamefont{Zhang}},
  \bibinfo{author}{\bibfnamefont{H.}~\bibnamefont{Ding}},
  \bibinfo{author}{\bibfnamefont{W.}~\bibnamefont{Li}},
  \bibinfo{author}{\bibfnamefont{L.}~\bibnamefont{Wang}},
  \bibinfo{author}{\bibfnamefont{K.}~\bibnamefont{He}},
  \bibinfo{author}{\bibfnamefont{C.-L.} \bibnamefont{Song}},
  \bibinfo{author}{\bibfnamefont{X.-C.} \bibnamefont{Ma}}, \bibnamefont{and}
  \bibinfo{author}{\bibfnamefont{Q.-K.} \bibnamefont{Xue}},
  \bibinfo{journal}{Science Bulletin} \textbf{\bibinfo{volume}{62}},
  \bibinfo{pages}{852 } (\bibinfo{year}{2017}), ISSN \bibinfo{issn}{2095-9273},
  \urlprefix\url{http://www.sciencedirect.com/science/article/pii/S2095927317302487}.

\bibitem[{\citenamefont{Sato and Ando}(2017)}]{Sato2017}
\bibinfo{author}{\bibfnamefont{M.}~\bibnamefont{Sato}} \bibnamefont{and}
  \bibinfo{author}{\bibfnamefont{Y.}~\bibnamefont{Ando}},
  \bibinfo{journal}{Reports on Progress in Physics}
  \textbf{\bibinfo{volume}{80}}, \bibinfo{pages}{076501}
  (\bibinfo{year}{2017}),
  \urlprefix\url{http://stacks.iop.org/0034-4885/80/i=7/a=076501}.

\bibitem[{\citenamefont{Elliott and Franz}(2015)}]{Elliott2015}
\bibinfo{author}{\bibfnamefont{S.~R.} \bibnamefont{Elliott}} \bibnamefont{and}
  \bibinfo{author}{\bibfnamefont{M.}~\bibnamefont{Franz}},
  \bibinfo{journal}{Rev. Mod. Phys.} \textbf{\bibinfo{volume}{87}},
  \bibinfo{pages}{137} (\bibinfo{year}{2015}),
  \urlprefix\url{https://link.aps.org/doi/10.1103/RevModPhys.87.137}.

\bibitem[{\citenamefont{Jackiw and Rebbi}(1976)}]{Jackiw76}
\bibinfo{author}{\bibfnamefont{R.}~\bibnamefont{Jackiw}} \bibnamefont{and}
  \bibinfo{author}{\bibfnamefont{C.}~\bibnamefont{Rebbi}},
  \bibinfo{journal}{Phys. Rev. D} \textbf{\bibinfo{volume}{13}},
  \bibinfo{pages}{3398} (\bibinfo{year}{1976}).

\bibitem[{\citenamefont{Takahashi}(1979)}]{Takahashi1979}
\bibinfo{author}{\bibfnamefont{K.}~\bibnamefont{Takahashi}},
  \bibinfo{journal}{Journal of Mathematical Physics}
  \textbf{\bibinfo{volume}{20}}, \bibinfo{pages}{1232} (\bibinfo{year}{1979}).

\bibitem[{\citenamefont{Bartsch and Ding}(2006)}]{Bartsch2006}
\bibinfo{author}{\bibfnamefont{T.}~\bibnamefont{Bartsch}} \bibnamefont{and}
  \bibinfo{author}{\bibfnamefont{Y.}~\bibnamefont{Ding}},
  \bibinfo{journal}{Journal of Differential Equations}
  \textbf{\bibinfo{volume}{226}}, \bibinfo{pages}{210} (\bibinfo{year}{2006}).

\bibitem[{\citenamefont{Haddad et~al.}(2015)\citenamefont{Haddad, Weaver, and
  Carr}}]{Haddad2015}
\bibinfo{author}{\bibfnamefont{L.~H.} \bibnamefont{Haddad}},
  \bibinfo{author}{\bibfnamefont{C.~M.} \bibnamefont{Weaver}},
  \bibnamefont{and} \bibinfo{author}{\bibfnamefont{L.~D.} \bibnamefont{Carr}},
  \bibinfo{journal}{New Journal of Physics} \textbf{\bibinfo{volume}{17}},
  \bibinfo{pages}{063033} (\bibinfo{year}{2015}).

\bibitem[{\citenamefont{Solnyshkov et~al.}(2016)\citenamefont{Solnyshkov,
  Nalitov, and Malpuech}}]{Solnyshkov2016}
\bibinfo{author}{\bibfnamefont{D.~D.} \bibnamefont{Solnyshkov}},
  \bibinfo{author}{\bibfnamefont{A.~V.} \bibnamefont{Nalitov}},
  \bibnamefont{and} \bibinfo{author}{\bibfnamefont{G.}~\bibnamefont{Malpuech}},
  \bibinfo{journal}{Phys. Rev. Lett.} \textbf{\bibinfo{volume}{116}},
  \bibinfo{pages}{046402} (\bibinfo{year}{2016}),
  \urlprefix\url{https://link.aps.org/doi/10.1103/PhysRevLett.116.046402}.

\bibitem[{\citenamefont{Gulevich et~al.}(2017)\citenamefont{Gulevich, Yudin,
  Skryabin, Iorsh, and Shelykh}}]{Skryabin2017}
\bibinfo{author}{\bibfnamefont{D.~R.} \bibnamefont{Gulevich}},
  \bibinfo{author}{\bibfnamefont{D.}~\bibnamefont{Yudin}},
  \bibinfo{author}{\bibfnamefont{D.~V.} \bibnamefont{Skryabin}},
  \bibinfo{author}{\bibfnamefont{I.~V.} \bibnamefont{Iorsh}}, \bibnamefont{and}
  \bibinfo{author}{\bibfnamefont{I.~A.} \bibnamefont{Shelykh}},
  \bibinfo{journal}{Scientific Reports} \textbf{\bibinfo{volume}{7}},
  \bibinfo{pages}{1780} (\bibinfo{year}{2017}).

\bibitem[{\citenamefont{Solnyshkov et~al.}(2017)\citenamefont{Solnyshkov, Bleu,
  Teklu, and Malpuech}}]{Solnyshkov2017}
\bibinfo{author}{\bibfnamefont{D.~D.} \bibnamefont{Solnyshkov}},
  \bibinfo{author}{\bibfnamefont{O.}~\bibnamefont{Bleu}},
  \bibinfo{author}{\bibfnamefont{B.}~\bibnamefont{Teklu}}, \bibnamefont{and}
  \bibinfo{author}{\bibfnamefont{G.}~\bibnamefont{Malpuech}},
  \bibinfo{journal}{Phys. Rev. Lett.} \textbf{\bibinfo{volume}{118}},
  \bibinfo{pages}{023901} (\bibinfo{year}{2017}),
  \urlprefix\url{https://link.aps.org/doi/10.1103/PhysRevLett.118.023901}.

\bibitem[{\citenamefont{Haddad and Carr}(2015)}]{Haddad2015b}
\bibinfo{author}{\bibfnamefont{L.~H.} \bibnamefont{Haddad}} \bibnamefont{and}
  \bibinfo{author}{\bibfnamefont{L.~D.} \bibnamefont{Carr}},
  \bibinfo{journal}{New Journal of Physics} \textbf{\bibinfo{volume}{17}},
  \bibinfo{pages}{113011} (\bibinfo{year}{2015}).

\bibitem[{\citenamefont{Peano et~al.}(2016)\citenamefont{Peano, Houde, Brendel,
  Marquardt, and Clerk}}]{peano2015topological}
\bibinfo{author}{\bibfnamefont{V.}~\bibnamefont{Peano}},
  \bibinfo{author}{\bibfnamefont{M.}~\bibnamefont{Houde}},
  \bibinfo{author}{\bibfnamefont{C.}~\bibnamefont{Brendel}},
  \bibinfo{author}{\bibfnamefont{F.}~\bibnamefont{Marquardt}},
  \bibnamefont{and} \bibinfo{author}{\bibfnamefont{A.~A.} \bibnamefont{Clerk}},
  \bibinfo{journal}{Nature communications} \textbf{\bibinfo{volume}{7}},
  \bibinfo{pages}{10779} (\bibinfo{year}{2016}),
  \urlprefix\url{http://dx.doi.org/10.1038/ncomms10779}.

\bibitem[{\citenamefont{Di~Liberto et~al.}(2016)\citenamefont{Di~Liberto,
  Hemmerich, and Smith}}]{di2016topological}
\bibinfo{author}{\bibfnamefont{M.}~\bibnamefont{Di~Liberto}},
  \bibinfo{author}{\bibfnamefont{A.}~\bibnamefont{Hemmerich}},
  \bibnamefont{and} \bibinfo{author}{\bibfnamefont{C.~M.} \bibnamefont{Smith}},
  \bibinfo{journal}{arXiv preprint arXiv:1604.06055}  (\bibinfo{year}{2016}).

\bibitem[{\citenamefont{Bardyn et~al.}(2016)\citenamefont{Bardyn, Karzig,
  Refael, and Liew}}]{Bardyn2016}
\bibinfo{author}{\bibfnamefont{C.-E.} \bibnamefont{Bardyn}},
  \bibinfo{author}{\bibfnamefont{T.}~\bibnamefont{Karzig}},
  \bibinfo{author}{\bibfnamefont{G.}~\bibnamefont{Refael}}, \bibnamefont{and}
  \bibinfo{author}{\bibfnamefont{T.~C.~H.} \bibnamefont{Liew}},
  \bibinfo{journal}{Phys. Rev. B} \textbf{\bibinfo{volume}{93}},
  \bibinfo{pages}{020502} (\bibinfo{year}{2016}),
  \urlprefix\url{http://link.aps.org/doi/10.1103/PhysRevB.93.020502}.

\bibitem[{\citenamefont{Bleu et~al.}(2016)\citenamefont{Bleu, Solnyshkov, and
  Malpuech}}]{Bleu2016}
\bibinfo{author}{\bibfnamefont{O.}~\bibnamefont{Bleu}},
  \bibinfo{author}{\bibfnamefont{D.~D.} \bibnamefont{Solnyshkov}},
  \bibnamefont{and} \bibinfo{author}{\bibfnamefont{G.}~\bibnamefont{Malpuech}},
  \bibinfo{journal}{Phys. Rev. B} \textbf{\bibinfo{volume}{93}},
  \bibinfo{pages}{085438} (\bibinfo{year}{2016}),
  \urlprefix\url{https://link.aps.org/doi/10.1103/PhysRevB.93.085438}.

\bibitem[{\citenamefont{Gulevich et~al.}(2016)\citenamefont{Gulevich, Skryabin,
  Alodjants, and Shelykh}}]{Gulevich2016}
\bibinfo{author}{\bibfnamefont{D.~R.} \bibnamefont{Gulevich}},
  \bibinfo{author}{\bibfnamefont{D.~V.} \bibnamefont{Skryabin}},
  \bibinfo{author}{\bibfnamefont{A.~P.} \bibnamefont{Alodjants}},
  \bibnamefont{and} \bibinfo{author}{\bibfnamefont{I.~A.}
  \bibnamefont{Shelykh}}, \bibinfo{journal}{Phys. Rev. B}
  \textbf{\bibinfo{volume}{94}}, \bibinfo{pages}{115407}
  (\bibinfo{year}{2016}),
  \urlprefix\url{https://link.aps.org/doi/10.1103/PhysRevB.94.115407}.

\bibitem[{\citenamefont{Bleu et~al.}(2017{\natexlab{b}})\citenamefont{Bleu,
  Solnyshkov, and Malpuech}}]{Bleu2017b}
\bibinfo{author}{\bibfnamefont{O.}~\bibnamefont{Bleu}},
  \bibinfo{author}{\bibfnamefont{D.~D.} \bibnamefont{Solnyshkov}},
  \bibnamefont{and} \bibinfo{author}{\bibfnamefont{G.}~\bibnamefont{Malpuech}},
  \bibinfo{journal}{Phys. Rev. B} \textbf{\bibinfo{volume}{95}},
  \bibinfo{pages}{115415} (\bibinfo{year}{2017}{\natexlab{b}}),
  \urlprefix\url{https://link.aps.org/doi/10.1103/PhysRevB.95.115415}.

\bibitem[{\citenamefont{Sigurdsson et~al.}(2017)\citenamefont{Sigurdsson, Li,
  and Liew}}]{Liew2017}
\bibinfo{author}{\bibfnamefont{H.}~\bibnamefont{Sigurdsson}},
  \bibinfo{author}{\bibfnamefont{G.}~\bibnamefont{Li}}, \bibnamefont{and}
  \bibinfo{author}{\bibfnamefont{T.~C.~H.} \bibnamefont{Liew}},
  \bibinfo{journal}{arXiv:1707.02776}  (\bibinfo{year}{2017}).

\bibitem[{\citenamefont{Jacqmin et~al.}(2014)\citenamefont{Jacqmin, Carusotto,
  Sagnes, Abbarchi, Solnyshkov, Malpuech, Galopin, Lema\^{\i}tre, Bloch, and
  Amo}}]{Jacqmin2014}
\bibinfo{author}{\bibfnamefont{T.}~\bibnamefont{Jacqmin}},
  \bibinfo{author}{\bibfnamefont{I.}~\bibnamefont{Carusotto}},
  \bibinfo{author}{\bibfnamefont{I.}~\bibnamefont{Sagnes}},
  \bibinfo{author}{\bibfnamefont{M.}~\bibnamefont{Abbarchi}},
  \bibinfo{author}{\bibfnamefont{D.~D.} \bibnamefont{Solnyshkov}},
  \bibinfo{author}{\bibfnamefont{G.}~\bibnamefont{Malpuech}},
  \bibinfo{author}{\bibfnamefont{E.}~\bibnamefont{Galopin}},
  \bibinfo{author}{\bibfnamefont{A.}~\bibnamefont{Lema\^{\i}tre}},
  \bibinfo{author}{\bibfnamefont{J.}~\bibnamefont{Bloch}}, \bibnamefont{and}
  \bibinfo{author}{\bibfnamefont{A.}~\bibnamefont{Amo}},
  \bibinfo{journal}{Phys. Rev. Lett.} \textbf{\bibinfo{volume}{112}},
  \bibinfo{pages}{116402} (\bibinfo{year}{2014}),
  \urlprefix\url{http://link.aps.org/doi/10.1103/PhysRevLett.112.116402}.

\bibitem[{\citenamefont{Soltan-Panahi et~al.}(2011)\citenamefont{Soltan-Panahi,
  Struck, Hauke, Bick, Plenkers, Meineke, Becker, Windpassinger, Lewenstein,
  and Sengstock}}]{Soltan2011}
\bibinfo{author}{\bibfnamefont{P.}~\bibnamefont{Soltan-Panahi}},
  \bibinfo{author}{\bibfnamefont{J.}~\bibnamefont{Struck}},
  \bibinfo{author}{\bibfnamefont{P.}~\bibnamefont{Hauke}},
  \bibinfo{author}{\bibfnamefont{A.}~\bibnamefont{Bick}},
  \bibinfo{author}{\bibfnamefont{W.}~\bibnamefont{Plenkers}},
  \bibinfo{author}{\bibfnamefont{G.}~\bibnamefont{Meineke}},
  \bibinfo{author}{\bibfnamefont{C.}~\bibnamefont{Becker}},
  \bibinfo{author}{\bibfnamefont{P.}~\bibnamefont{Windpassinger}},
  \bibinfo{author}{\bibfnamefont{M.}~\bibnamefont{Lewenstein}},
  \bibnamefont{and}
  \bibinfo{author}{\bibfnamefont{K.}~\bibnamefont{Sengstock}},
  \bibinfo{journal}{Nature Physics} \textbf{\bibinfo{volume}{7}},
  \bibinfo{pages}{434} (\bibinfo{year}{2011}).

\bibitem[{\citenamefont{Martin et~al.}(2008)\citenamefont{Martin, Blanter, and
  Morpurgo}}]{Martin2008}
\bibinfo{author}{\bibfnamefont{I.}~\bibnamefont{Martin}},
  \bibinfo{author}{\bibfnamefont{Y.~M.} \bibnamefont{Blanter}},
  \bibnamefont{and} \bibinfo{author}{\bibfnamefont{A.~F.}
  \bibnamefont{Morpurgo}}, \bibinfo{journal}{Phys. Rev. Lett.}
  \textbf{\bibinfo{volume}{100}}, \bibinfo{pages}{036804}
  (\bibinfo{year}{2008}),
  \urlprefix\url{http://link.aps.org/doi/10.1103/PhysRevLett.100.036804}.

\bibitem[{sup()}]{suppl}
\bibinfo{note}{See Supplemental Material at [URL will be inserted by
  publisher].}

\bibitem[{\citenamefont{Xiao et~al.}(2012)\citenamefont{Xiao, Liu, Feng, Xu,
  and Yao}}]{Xiao2012}
\bibinfo{author}{\bibfnamefont{D.}~\bibnamefont{Xiao}},
  \bibinfo{author}{\bibfnamefont{G.-B.} \bibnamefont{Liu}},
  \bibinfo{author}{\bibfnamefont{W.}~\bibnamefont{Feng}},
  \bibinfo{author}{\bibfnamefont{X.}~\bibnamefont{Xu}}, \bibnamefont{and}
  \bibinfo{author}{\bibfnamefont{W.}~\bibnamefont{Yao}},
  \bibinfo{journal}{Phys. Rev. Lett.} \textbf{\bibinfo{volume}{108}},
  \bibinfo{pages}{196802} (\bibinfo{year}{2012}),
  \urlprefix\url{http://link.aps.org/doi/10.1103/PhysRevLett.108.196802}.

\bibitem[{\citenamefont{Pitaevskii}(1959)}]{Pitaevskii58}
\bibinfo{author}{\bibfnamefont{L.~P.} \bibnamefont{Pitaevskii}},
  \bibinfo{journal}{Sov. Phys. JETP} \textbf{\bibinfo{volume}{35}},
  \bibinfo{pages}{282} (\bibinfo{year}{1959}).

\bibitem[{\citenamefont{Tanese et~al.}(2013)\citenamefont{Tanese, Flayac,
  Solnyshkov, Amo, Lemaitre, Galopin, Braive, Senellart, Sagnes, Malpuech
  et~al.}}]{Tanese2013}
\bibinfo{author}{\bibfnamefont{D.}~\bibnamefont{Tanese}},
  \bibinfo{author}{\bibfnamefont{H.}~\bibnamefont{Flayac}},
  \bibinfo{author}{\bibfnamefont{D.}~\bibnamefont{Solnyshkov}},
  \bibinfo{author}{\bibfnamefont{A.}~\bibnamefont{Amo}},
  \bibinfo{author}{\bibfnamefont{A.}~\bibnamefont{Lemaitre}},
  \bibinfo{author}{\bibfnamefont{E.}~\bibnamefont{Galopin}},
  \bibinfo{author}{\bibfnamefont{R.}~\bibnamefont{Braive}},
  \bibinfo{author}{\bibfnamefont{P.}~\bibnamefont{Senellart}},
  \bibinfo{author}{\bibfnamefont{I.}~\bibnamefont{Sagnes}},
  \bibinfo{author}{\bibfnamefont{G.}~\bibnamefont{Malpuech}},
  \bibnamefont{et~al.}, \bibinfo{journal}{Nature Comm.}
  \textbf{\bibinfo{volume}{4}}, \bibinfo{pages}{1749} (\bibinfo{year}{2013}).

\end{thebibliography}

\section{Supplemental material}
In this supplemental material, we present additional details on the derivation of results of the main text. We discuss the winding-valley coupling and the velocity of a vortex at an interface. Finally, we comment the supplemental video files.

\subsection{Vortex-valley coupling}
The calculation of the Fourier transform $\widetilde{\psi}(\mathbf{k})$ from the main text is carried out as follows. In the TB approximation, $\psi(\mathbf{r})$ is defined only in discrete points in space, and the integration is replaced by summation. Studying the core only, we take into account only the 3 atoms of the $A$ type of the central hexagon. 

This gives the following sum:
\begin{eqnarray}
\widetilde{\psi}_p \left( \mathbf{k} \right) &=& {e^{i\left( {0 - \left( {{k_x},{k_y}} \right)\left( {0,0} \right)} \right)}} + {e^{i\left( {\frac{{2\pi }}{3}p - \left( {{k_x},{k_y}} \right)\left( {\frac{{3a}}{2},\frac{{a\sqrt 3 }}{2}} \right)} \right)}}\\ \nonumber
 &+& {e^{i\left( {\frac{{4\pi }}{3}p - \left( {{k_x},{k_y}} \right)\left( {0,a\sqrt 3 } \right)} \right)}} 
\end{eqnarray}
 where $p=\pm 1$ is the vortex winding.
This expression can be rewritten as
\begin{eqnarray}
\widetilde{\psi}_p \left( \mathbf{k} \right) &=& 1 + {e^{i\left( {\frac{{2\pi }}{3}p - \frac{3}{2}a{k_x} - \frac{{\sqrt 3 }}{2}a{k_y}} \right)}}\\ \nonumber
 &+& {e^{i\left( {\frac{{4\pi }}{3}p - \sqrt 3 a{k_y}} \right)}}
\end{eqnarray}
To simplify the expressions, let us define the arguments of the two exponents as separate variables:
\begin{equation}
\eta_p={\frac{{2\pi }}{3}p - \frac{3}{2}a{k_x} - \frac{{\sqrt 3 }}{2}a{k_y}}
\end{equation}
\begin{equation}
\zeta_p={\frac{{4\pi }}{3}p - \sqrt 3 a{k_y}}
\end{equation}
We can then find the position of the maximal probability density in the reciprocal space $|\psi(\mathbf{k})|^2$, which writes (by separating the real and imaginary parts):
\begin{eqnarray}
{\left| {\widetilde{\psi}_p \left( \mathbf{k} \right)} \right|^2} &=& 1 + {\cos ^2}\eta_p  + {\cos ^2}\zeta_p  + 2\cos \eta_p  + 2\cos \zeta_p \\ \nonumber
 &+& 2\cos \eta_p \cos \zeta_p  + {\sin ^2}\eta_p  + {\sin ^2}\zeta_p  + 2\sin \eta_p \sin \zeta_p 
\end{eqnarray}
which can be simplified to
\begin{equation}
\left|\widetilde{\psi}_p\left(\mathbf{k}\right)\right|^2=3+2\left(\cos\eta_p+\cos\zeta_p+2\cos\eta_p\cos\zeta_p\right)
\end{equation}

The maximal value of this expression is achieved when both $\eta_p=2\pi \nu$ and $\zeta_p=2\pi \mu$, where $\nu$ and $\mu$ are integer numbers. From the latter, taking for example $\nu=0$, it is easy to obtain, for $p=1$, $k_y=K$ (where $K=4\pi/3\sqrt{3}a$), and $k_x=0$, and for $p=-1$, $k_y=-K$ and $k_x=0$.

 \subsection{Vortex velocity}
\begin{figure}[bp]
 \begin{center}
 \includegraphics[scale=1.3]{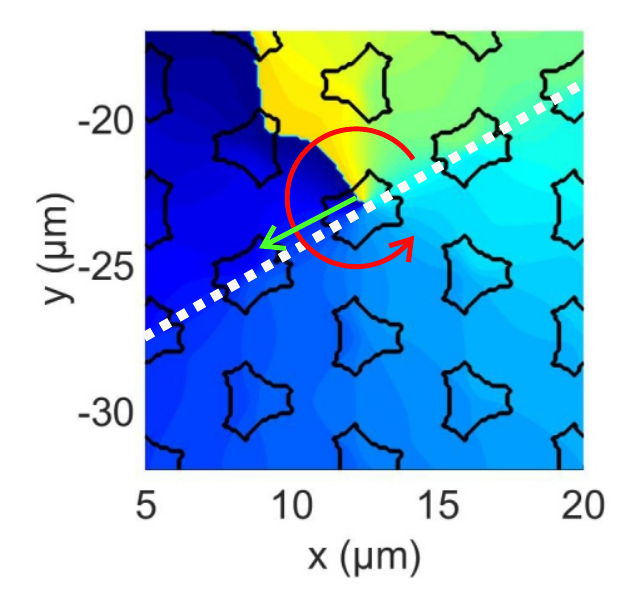}
 \caption{\label{figCore} Contour plot of the potential (black line) and the phase of the vortex (in color). Red arrow shows the rotation direction, green arrow shows the propagation direction of the vortex.}
  \end{center}
 \end{figure}

We have studied how the vortex velocity depends on the parameters of the system in order to check that the propagation along the interface is not linked with the well-known vortex rolling effect. 
First, let us see that the vortex really follows the interface, and its core is located exactly  within the unit cell, which separates the two inverted materials. Figure S\ref{figCore} shows a snapshot of the phase of the wavefunction with a vortex. A $2\pi$ phase jump line is clearly visible, and the core of the vortex is located at the end of this line. The rotation direction of the vortex is shown with a red arrow, and the green arrow indicates the propagation direction of the vortex along the interface (white dashed line). We see that the edge of the phase jump line is within the unit cell located at the interface.

\begin{figure}[tbp]
 \begin{center}
 \includegraphics[scale=0.33]{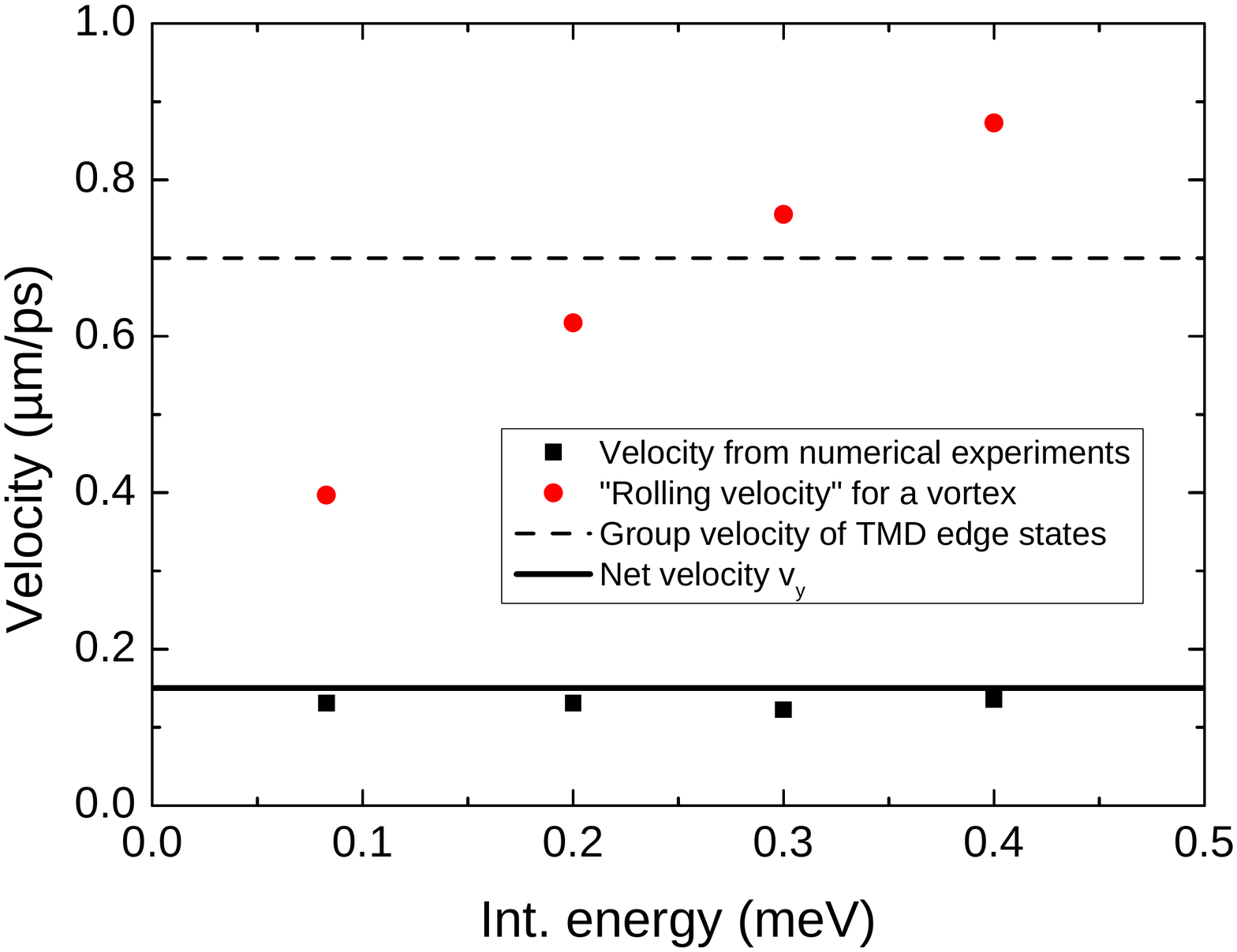}
 \caption{\label{figVvel} Vortex velocity from numerical calculations and its estimation by different models.}
  \end{center}
 \end{figure}

One might think that the vortex is simply rolling along the interface, like a wheel, converting rotation into propagation. The characteristic distance at which the density can vary in the condensate is given by the healing length $\xi$
and therefore the center of the vortex in this "rolling wheel" image has to be located at a distance $\xi$ from the wall, which allows to find the speed of rotation of the particles where they meet with the wall (and therefore the vortex propagation speed) using the expression
\begin{equation}
v=\frac{\hbar}{m}\frac{1}{r}
\end{equation}
where one takes $r=\xi$, which gives simply that the vortex propagates with a velocity roughly equal to the speed of sound in the condensate $v=\sqrt{\alpha n/m}$. In this model, one could therefore expect a pronounced dependence of the vortex propagation velocity on the particle density.
Another alternative could be that the vortex simply propagates with the group velocity of linear states at the interface, which can be calculated from the dispersion, as discussed in the main text. Figure S\ref{figVvel} compares the predictions of these models as a function of interaction energy $\alpha n$ with numerical results (black squares). Clearly, the simple predictions of the two naive models (red circles for rolling effect and black dashed line for the linear group velocity) strongly deviate from numerics. The model of the rolling wheel (red dots) predicts a dependence on the density which is not observed at all (the interaction energy changes by a factor 5, and there is no significant change of the vortex velocity). The group velocity of the interface states strongly overestimates the real vortex propagation speed (also by a factor 5).

To calculate the vortex velocity, we analyze the currents that take place within its core (concentrated in a given valley because of winding-valley coupling). In the bulk, the valley states are not propagating, but rotating, because the 3 quantum-mechanical current terms between the 3 pillars of the same type which have different phases ($0$, $2\pi/3$, $4\pi/3$) exactly compensate each other, as these are three identical vectors rotated at 120 degrees. Indeed,
\[j = \frac{{n\hbar }}{m}\nabla \varphi \]
where $n$ is the particle density, and therefore, to calculate current in the tight-binding approach we need to consider only pillars with nonzero density and take into account the phase difference between each pair.

At the interface the situation changes, as can be seen in Fig.~4(a) of the main text. The $A$ pillars on the left of the interface are not large pillars (with lower energy) but small pillars (with higher energy), and therefore, the 3 current terms (blue arrows) do not have the same prefactor. The phase differences are the same, but the density on the pillars on the left of the interface is smaller (it is not zero as it would be in the bulk, because the presence of the interface mixes the Bloch states), and therefore the current term marked as a dashed line has a smaller magnitude than the other two. This results in a net current pointing upwards, and this is what leads to the propagative nature of the interface states.

The total current reads
\[{\bf{j}} = {{\bf{j}}_1} + {{\bf{j}}_2} + {{\bf{j}}_3}.\]
Assuming that the density on the $A$ pillars on the right of the interface is $n$ and the density on the $A$ pillars on the left of the interface is $n'$, we can write the magnitude of the current terms as:
\[{j_{1,2}} = \frac{n+n'}{2}\frac{{\hbar }}{m}\frac{{2\pi }}{{3\sqrt 3 a}}\]
and
\[{j_{3}} = \frac{n'\hbar }{m}\frac{{2\pi }}{{3\sqrt 3 a}}\]
The orientation of the vectors makes that the $X$ projection of $j_3$ is $0$, while the $X$ projections of $j_1$ and $j_2$ are opposite, and so they compensate each other. The $Y$ projections give:
\[{j_Y} = \frac{1}{2}\left(j_1+j_2\right)  - {j_3}\]
which finally gives
\[{j_Y} = \frac{{n - n'}}{2}\frac{\hbar }{m}\frac{{2\pi }}{{3\sqrt 3 a}}\]
Without the interface, $n=n'$ and $\mathbf{j}=0$, as expected. The presence of the interface makes $n'<n$. If we consider an isolated problem of two pillars with coupling $J$ and energy splitting $\Delta$ (which determines the gap in the bulk TMD analog), we can estimate $n'$ as
\begin{equation}
n' = \frac{2n}{{1 + {{\left( {\Delta  + \sqrt {{\Delta ^2} + 4{J^2}} } \right)}^2}/4{J^2}}}
\label{nprimeeq}
\end{equation}
which finally gives the expression for the group velocity of the main text, because $2\pi\hbar/m/3\sqrt{3}a$ is simply an estimate of the group velocity $v_g$ in terms of the tight-binding parameters.

We can also calculate an approximated expression, assuming that $\Delta\ll J$, 
\begin{equation}
n' = n\left( {1 - \frac{\Delta }{{2J}}} \right)
\label{nprime}
\end{equation}
which gives for the net velocity along the interface
\begin{equation}
{v_Y} \approx \frac{\Delta }{{2J}}\frac{\hbar }{m}\frac{{2\pi }}{{3\sqrt 3 a}}
\label{eqVy}
\end{equation}
The corresponding calculated velocity shown in Fig. S2 by a solid black line corresponds well to the numerical results, contrary to the predictions of the simple models.

In the opposite limit of very large $\Delta$,
\[n' = n\frac{{2{J^2}}}{{{\Delta ^2}}}\]
and
\begin{equation}
{v_Y} \approx \left( {1 - \frac{{2{J^2}}}{{{\Delta ^2}}}} \right)\frac{\hbar }{m}\frac{{2\pi }}{{3\sqrt 3 a}}
\label{eqLim}
\end{equation}

This expression also increases with the increase of $\Delta$. It is interesting to see that this expression is bounded from above by a limiting value, which cannot be exceeded by changing $\Delta$ (but only by changing $J$, which affects $m$).

\section{Supplemental video}
In the supplemental video file \textsf{vortexdefect.avi} (also available at \url{https://www.youtube.com/watch?v=PNsDF5xUvH4}), we show the temporal evolution of the spatial density distribution of the condensate $|\psi(\mathbf{r},t)|^2$, obtained by direct solution of the Gross-Pitaevskii equation with $U(\mathbf{r})$ being the lattice potential, without the tight-binding approximation. The snapshots from this movie are shown in Fig. 3 of the main text. The vortex is attached to one side of the interface and propagates along it, passing around two corners and a defect.

A second movie \textsf{linwp.avi} (also available at \url{https://www.youtube.com/watch?v=M7nbL5i9l44}) demonstrates that a linear Gauss-Laguerre wavepacket with a non-zero angular momentum does not at all exhibit the same behavior as the vortex in an interacting condensate: the wavepacket is unstable and expands rapidly, preventing the observer to keep trace of the propagation of its center. The features of the interacting BEC maintaining the vortex are therefore crucial for the results obtained in the main text.

\end{document}